# Site-engineered ferromagnetism in Ca and Cr co-substituted Bismuth Ferrite Nanoparticles

**Authors :** Mehedi Hasan Prince [a,b †], Abrar Daiyan [a †], Troyee Mitra Aishi [a,c †], Anika Rahman Riya [a], Md. Fakhrul Islam [d], Md. Abdullah Zubair [d], Takian Fakhrul [a,*]

**Affiliations** :

[a] Department of Materials and Metallurgical Engineering, Bangladesh University of Engineering and Technology (BUET), Dhaka, 1000, Bangladesh

[b] Department of Materials Science & Engineering, Rensselaer Polytechnic Institute (RPI), NY, 12180, USA

[c] Department of Chemical Engineering, Northeastern University, MA, 02115, USA

[d] Department of Nanomaterials and Ceramic Engineering, Bangladesh University of Engineering and Technology (BUET), Dhaka, 1000, Bangladesh

* Corresponding author. Email: takianf@mme.buet.ac.bd

† Mehedi Hasan Prince, Abrar Daiyan, Troyee Mitra Aishi contributed equally to this work.

abstract
**Abstract**

Multiferroic perovskites that exhibit room temperature magnetization and polarization have immense potential in the next generation of magneto-electric and spintronic memory devices. In this work, the magnetic and ferroelectric properties of Bismuth Ferrite, $BiFeO_3$ (BFO) nanoparticles (NPs) were enhanced through simultaneous A- and B-site Ca and Cr co-substitution. Novel compositions of $Bi_{0.97}Ca_{0.03}Cr_xFe_{1-x}O_3$ (x=0, 0.01, 0.03, 0.05) were synthesized using the sol-gel route and annealed at 550°C. Rietveld Refinement of XRD patterns confirmed high phase purity, while SEM analysis revealed a decreasing trend in average particle size with increasing dopant concentration. Hysteresis loops showed enhanced magnetic properties as particle size approached the spin cycloid wavelength (~62 nm), disrupting the intrinsic antiferromagnetic ordering of BFO. Moreover, the presence of exchange bias in the NPs was linked to the formation of core-shell structure. Temperature dependent magnetization studies showed an increase in Néel temperature upon Ca substitution. XPS analysis confirmed that $Bi_{0.97}Ca_{0.03}FeO_3$ samples exhibited the highest oxygen vacancy concentration, while $Fe^{3+}$ remained the dominant oxidation state across all compositions. Ferroelectric polarization loop measurements showed enhanced remanent polarization in doped samples, with leakage linked to oxygen vacancies and extrinsic microstructural effects.

# 1 Introduction

Materials that simultaneously exhibit two or more fundamental ferroic orders, such as ferromagnetism, ferroelectricity, and ferroelasticity, within a single phase are known as multiferroic materials [1]. Their multifunctional nature, facilitated by the coupling between these orders, has garnered interest from researchers due to their applications in ferroelectric random access memory, magnetoelectric sensors and actuators, spintronics, photovoltaics, magneto-optical devices, tunable photonic devices etc [2–7]. Among single-phase multiferroics, $BiFeO_3$ (BFO) stands out as one of the few compounds exhibiting both ferroelectric and antiferromagnetic ordering at room temperature [1]. BFO has a distorted rhombohedral perovskite structure classified under the R3c space group with an anti-ferromagnetic Néel temperature ($T_N$) of 370 °C and ferroelectric Curie temperature ($T_C$) of 830 °C. Regardless of its potential commercial applications, limitations such as high leakage current, poor magnetoelectric coupling, secondary phase formation, and poor phase stability have hindered its full-scale integration into devices [8].

In its bulk form, BFO exhibits a spatially modulated spin cycloid with a wavelength of 62 nm, which dictates its G-type antiferromagnetic behavior. This helical structure leads to spin cancellation, reducing the net magnetization and limiting BFO's potential for integration into magnetoelectrically coupled devices [9]. Furthermore, the fabrication of pure $BiFeO_3$ remains a challenge, as it is generally accompanied by secondary phases such as $Bi_2O_3$, $Bi_2Fe_4O_9$, $Bi_{25}FeO_{39}$, etc [10,11]. However, significant studies have been conducted on the various synthesis routes such as sol-gel, solid state, rapid liquid phase sintering, hydrothermal, co-precipitation methods in order to maximize purity and enhance multiferroic properties [12–15]. Among these techniques, sol-gel has proven to be an eco-friendly and cost-effective process for synthesizing BFO nanoparticles [16,17].

An effective approach to addressing the limitations of bulk BFO is through nanoscale particle size reduction and site substitution [18]. The key to unlocking ferromagnetism in BFO is through unwinding the spin-cycloid, thereby generating a net magnetic moment in the process [19]. Since BFO is classified as a Type-I multiferroic, with ferroelectricity and ferromagnetism originating from separate sources, researchers have made significant strides in property enhancement through site engineering of A-site $Bi^{3+}$ and B-site $Fe^{3+}$ with rare earth, transition metal and alkaline earth ions [20–22]. Gitanjali et al. observed that substituting $Bi^{3+}$ with divalent $Ca^{2+}$ in the A-site resulted in a reduction in particle size, significantly improving the material's magnetic properties through spin cycloid suppression. They also observed improved ferroelectric properties in contrast to bulk BFO through leakage current reduction [23]. Further research by Bhushan et al. indicates that among A-site alkaline earth substituents, $Ca^{2+}$ resulted in the highest saturation and remnant magnetization. In addition to the breakdown of spin structure, this enhancement can be attributed to uncompensated spins arising from oxygen vacancies created for charge compensation [31]. Conversely, researchers have found that substituting B-site $Fe^{3+}$ with transition metal ions not only reduces leakage current but also enhances magnetism [24]. Replacing $Fe^{3+}$ with $Cr^{3+}$ is an effective way of enhancing the multiferroic properties of BFO while maintaining charge neutrality, suppressing the formation of oxygen vacancies [25]. Kumar et al. reported decreased secondary phase formation, reduced band gap, and increased ferromagnetic properties upon Cr doping [26]. Simultaneous substitution of both A and B site cations is key to improving the weak magneto-electric coupling present in Type-I multiferroics [27,28]. This investigation aims to bridge the existing gap in the literature by exploring the impact of both A-site ($Ca^{2+}$) and B-site ($Cr^{3+}$) substitution on the multiferroic properties of BFO nanoparticles.

In this study, pure, Ca doped and Ca-Cr co-substituted BFO NPs of compositions $BiFeO_3$ and $Bi_{0.97}Ca_{0.03}Fe_{1-x}Cr_xO_3$ (x=0.0,0.03,0.05) were synthesized using the sol-gel route. Using a range of characterization techniques, including X-Ray Diffraction (XRD), Scanning Electron Microscopy (SEM), X-ray Photoelectron Spectroscopy (XPS), Vibrating Sample Magnetometry (VSM), and Polarization-Electric Field (P-E) loop measurements, we explored the effect of simultaneous A-site and B-site substitution on the structural, magnetic and ferroelectric properties of the novel BFO nanoparticle compositions.

## 2 Experimental section

### 2.1 Materials

Analytical grade (AR) Bismuth Nitrate pentahydrate [$Bi(NO_3)_3 \cdot 5H_2O$, ≥99.0%], Iron nitrate nonahydrate [$Fe(NO_3)_3 \cdot 9H_2O$, ≥98.0%], Calcium Nitrate tetrahydrate [$Ca(NO_3)_2 \cdot 4H_2O$, ≥98.0%], Chromium Nitrate nonahydrate [$Cr(NO_3)_3 \cdot 9H_2O$, ≥98.0%] were used as precursors. Citric acid ($C_6H_8O_7$), ethylene glycol ($C_2H_6O_2$) and DI (de-ionized) water were employed as chelating agent, stabilizing agent and solvent respectively. All chemicals were supplied by Merck.

### 2.2 Preparation of $BiFeO_3$ nanoparticles

Modified sol-gel method was used for the synthesis of pure and Ca-Cr co-doped BFO NPs with compositions $BiFeO_3$ and $Bi_{0.97}Ca_{0.03}Fe_{1-x}Cr_xO_3$ [(BCFO; x=0), (BCFOCr-1; x= 0.01), (BCFOCr-3; x= 0.03), (BCFOCr-5; x= 0.05)]. **Figure 1** schematically depicts processing of the desired BFO based nanoparticles. To produce 2 gm sample, stoichiometric amount of the precursor salts and citric acid were dissolved in 400 ml DI water to form a solution. Citric acid was added as a chelating agent to bind all the metal cations by forming a complex compound with the metals and the molar ratio of iron and bismuth nitrate to citric acid was kept at 1:1. The solution was stirred on a hot plate at a temperature of 90–95°C for 30 minutes using a magnetic stirrer at 400 rpm. This was followed by the addition of 10 ml Ethylene Glycol to stabilize the gel by forming polymeric chains with the metal cations. The solution was then heated at 75-85°C while stirring at 350 rpm for about 4 hours, or until a dark brown viscous gel was obtained. The gel was dried at 110°C for about 24 hours and crushed into a powder using an agate mortar and pestle. To obtain the desired nanoparticles, the powders were then annealed at 550°C for about 2 hours with a heating rate of 5°C/min.

[Insert Figure 1 around here]

*2.3  Characterization*

Room Temperature (RT) phase and structure analysis of the annealed nanoparticles were carried out with an X-Ray Diffractometer (Empyrean PANalytical) equipped with a hybrid PIXcel$^{3D}$ detector, utilizing Cu X-ray source of wavelengths K$\alpha_1$ = 1.540598 Å and K$\alpha_2$ = 1.544426 Å. The X-ray source was operated at 45 kV and 40 mA. A zero-background silicon holder containing the sample was rotated at a speed of 60 rpm by a reflection-transmission spinner module to get an even area of irradiation both at low and high diffraction angles. The scanning speed was 0.041683°/sec, and the diffraction patterns were collected within a 2θ range of 20-80°. Xpert High Score plus was used for Rietveld refinement to extract the crystallographic information from the XRD patterns. To examine the microstructure and morphology of the samples, Field Emission Scanning Electron Microscope (FESEM: JEOL JSM 7600F) equipped with an EDX system was utilized. The room temperature magnetic property was measured by using Vibrating sample magnetometer (VSM: PPMS DynaCool, Quantum Design; NORTH AMERICA) by applying a magnetic field ranging from -20 kOe to +20 kOe. For electrical property measurement, 600 mg of NPs were pressed into pellets through an automated hydraulic press, applying a load of 3 tons for 60 seconds. The pellets were consolidated at a sintering temperature as low as 450°C for 2 hours, to restrict any additional particle growth and were subsequently coated with silver paste on both sides. Precision Materials Analyzer (Radiant

Technologies, Inc.: P-PMF, PMF0215-377) equipped with a 10 kV HVI-SC interface, signal amplifier and VISION software were used to get RT P-E loops.

## 3 Result and Discussion

### 3.1 Structural analysis

The measured RT XRD patterns of pure and Ca-Cr co-substituted BFO nanoparticles (NPs) together with the corresponding calculated patterns after Rietveld refinement, and the difference between the calculated and observed patterns are shown in **Figure 2.** The unit cell of each composition is also shown alongside the XRD patterns. All of the indexed peaks together with the characteristic doublet peaks of BFO, (104) and (110) at around 31.6° and 32°, confirmed the presence of rhombohedrally distorted perovskite structure with R3c space group having hexagonal symmetry [29]. The minor peaks observed at approximately 27° in the diffraction patterns of all nanoparticle samples (indicated by "S") indicate the presence of $Bi_2O_3$ while peaks, marked as "B," appearing exclusively in the BCFO sample, corresponds to the $Bi_2Fe_4O_9$ phase. These findings align with previously reported literature [25,30].

[Figure 2 around here]

**Table 1** shows the relative amounts and crystallographic symmetry information of different phases, together with unit cell parameters of the primary phase extracted from the Rietveld structural refinement of the XRD patterns. The goodness-of-fit ($\chi^2$) values for refinement fitting, ranging from 2.04 to 2.42, confirm the reliability of the extracted data. XRD analysis also shows an angular shift in both (104) and (110) peaks to higher diffraction angles with dopant incorporation, indicating that the lattice parameters of the NPs have decreased and subsequently lead to a gradual decrease in unit cell volume. Such change in lattice dimensions can be attributed to the fact that the effective ionic radii of the dopants $Ca^{2+}$ (1.00Å) and $Cr^{3+}$ (0.615 Å) are smaller than $Bi^{3+}$ (1.03 Å) and $Fe^{3+}$ (0.645 Å) [31,32]. According to Vegard's law [33], the incorporation of $Ca^{2+}$ and $Cr^{3+}$, having smaller ionic radii, in the A-site and B-site respectively decreases the unit cell parameters. Moreover, incorporation of $Ca^{2+}$ ion in $Bi^{3+}$ site is expected to cause oxygen vacancies ($V_O^{\bullet\bullet}$) in order to maintain charge neutrality, as depicted by the Kröger-Vink notation given below:

$$2Ca^{2+} \xrightarrow{Bi_2O_3} 2Ca'_{Bi} + 2O_O^x + V_O^{\bullet\bullet} \qquad (1)$$

Formation of oxygen vacancies causes neighboring atoms to move closer, leading to a reduction in the lattice spacing and a decrease in the unit cell volume.

**Table 1**

Structural data, phase percent, reliability factor, Angular distortion of pure and Ca, Cr co-doped Bismuth Ferrite NPs. *PP: Primary phase; *SP: second phase

| Sample | Phase present in volume % | Unit cell parameter of PP* | Rietveld factors |
|---|---|---|---|
| BFO | [PP] | a = b = 5.5791 Å | $R_p$ = 4.89236 |
|  | R3c (93.8%) | c = 13.868 Å | $R_{wp}$ = 6.57407 |
|  | $Bi_2O_3$ (SP) | V = 373.83 Å$^3$ | $R_{exp}$ = 3.06189 |
|  | P-4b2 (6.2%) |  | $\chi^2$ = 2.14 |
| BCFO | [PP] | a = b = 5.5797 Å | $R_p$ = 5.35179 |
|  | R3c (80%) | c = 13.860 Å | $R_{wp}$ = 6.97361 |
|  | $Bi_2O_3$ (SP) | V = 373.70 Å$^3$ | $R_{exp}$ = 3.41225 |
|  | P-4b2 (20%) |  | $\chi^2$ = 2.04370 |
| BCFOCr-1 | [PP] | a = b = 5.5797 Å | $R_p$ = 6.30904 |
|  | R3c (94.6%) | c = 13.857 Å | $R_{wp}$ = 8.45131 |
|  | $Bi_2O_3$ (SP) | V = 373.62 Å$^3$ | $R_{exp}$ = 3.48228 |
|  | P-421c (5.4%) |  | $\chi^2$ = 2.42694 |
| BCFOCr-3 | [PP] | a = b = 5.5783 Å | $R_p$ = 5.33679 |
|  | R3c (91.5%) | c = 13.853 Å | $R_{wp}$ = 7.17794 |
|  | $Bi_2O_3$ (SP) | V = 373.30 Å$^3$ | $R_{exp}$ = 3.45207 |
|  | P-421c (8.5%) |  | $\chi^2$ = 2.07931 |
| BCFOCr-5 | [PP] | a = b = 5.5764 Å | $R_p$ = 5.26068 |
|  | R3c (89.9%) | c = 13.852 Å | $R_{wp}$ = 6.82449 |
|  | $Bi_2O_3$ (SP) | V = 373.04 Å$^3$ | $R_{exp}$ = 3.51606 |
|  | P-421c (10.1%) |  | $\chi^2$ = 2.07931 |

Following the structure-less Rietveld analysis, the observed XRD patterns best fit the rhombohedral symmetry, indicating no evidence of major crystallographic phase change with dopant incorporation.

Crystallite size and micro strain were calculated using the formulae given below:

$$d = \frac{k\gamma}{\beta \cos \theta} \qquad (3)$$

$$\varepsilon = \frac{\beta}{4 \tan \theta} \qquad (4)$$

Where, k~0.9 is the shape factor, λ=1.5418 Å is the wavelength of Cu K$\alpha_{1,2}$ radiation, β is the instrument corrected Full width half maximum (FWHM) (in radians) or structural broadening, θ is the bragg angle (in radians). For crystallite size and strain calculation adopting the above equations, the structural broadening must correspond to a low angle, low index and high intensity distinct (hkl) reflection, free from influence of preferred orientation. Thus, for determination of β, (012) reflection has been chosen.

The maximum average crystallite size was observed for BFO (38.3 nm) and minimum was observed for BCFOCr-5 (21.6 nm). As the doping concentration increases, the lattice micro-strain in the nanoparticles follows an increasing trend, reaching a maximum for BCFOCr-5. In addition, the dislocation density was calculated using the following equation:

$$\delta = \frac{1}{d^2} \qquad (5)$$

Calculated crystallite sizes, lattice strains and dislocation densities are presented in **Table 2**.

**Table 2**

Crystallite size, Dislocation density, Lattice strain

| Composition | Crystallite size, d (nm) | Dislocation density, δ (nm$^{-2}$) (×10$^{-4}$) | Lattice strain, ε (×10$^{-3}$) |
|---|---|---|---|
| BFO | 38.3 | 0.068 | 4.87 |
| BCFO | 27.7 | 0.130 | 6.72 |
| BCFOCr-1 | 24.18 | 0.172 | 7.75 |
| BCFOCr-3 | 22.1 | 0.205 | 8.44 |
| BCFOCr-5 | 21.6 | 0.214 | 9.43 |

*3.2 Morphological analysis*

The microstructural images of the BFO, BCFO, BCFOCr-1, BCFOCr-3, and BCFOCr-5 NPs are shown in **Figure 3** along with the particle size distribution. Morphological analysis revealed that the average particle size was under 100 nm for all the compositions. Some amount of surface porosities were observed in BFO sample, which could be attributed to the difference in the diffusion rate of the constituting elements (Bi, Fe & O) of BFO [49]. In addition, uniformly distributed regular shaped particles were observed in the BFO sample. In case of BCFO, the particle size decreased from 83 nm to 74 nm along with increased shape irregularity. The observed non-uniform particle size distribution, greater degree of particle agglomeration and higher surface porosities upon Ca$^{2+}$ substitution were an expected outcome that resulted from the generation of oxygen vacancies [34]. With increased Cr concentration, the particle size further decreased, as depicted in **Figure 3**, which could be attributed to the introduction of additional nucleation sites generated by $Cr^{3+}$ addition. Moreover, dopants may segregate as impurities or defects that restrict crystal growth through interfacial pinning. This disruption in the regular atomic arrangement makes interfacial migration an energetically unfavorable phenomenon, thereby leading to the formation of smaller crystallites.

[ Insert Figure 3 around here]

All of the NPs' crystallite size (calculated form the Scherrer's formula) were less than the actual average particle size obtained from SEM analysis. In the sol-gel process, nanocrystals are shielded by the complexing ligands, preventing particle agglomeration and leading to the formation of very fine nanocrystals. However, the citrate complexing agent evaporates at around 300°C. Therefore, at annealing temperature of 550°C, bare multiferroic nanocrystals with high surface energies remain, leading to the formation of inter-particle necks through solid-state diffusion. This process facilitates partial agglomeration and particle growth as the system seeks to minimize surface energy [35].

*3.3   Magnetic Property Analysis*

**Figure 4** shows the room temperature M-H curves for pure and co-doped BFO nanoparticles measured using VSM in an external magnetic field ranging from -20 kOe to +20 kOe. Good hysteresis behavior was observed in the M-H curves and the values of saturation magnetization ($M_s$), remnant magnetization ($M_r$) and coercivity ($H_c$) were obtained, as listed in **Table 3**. With Cr substitution in $Bi_{0.97}Ca_{0.03}Cr_xFe_{1-x}O_3$, we observed an increase in ($M_s$), ($M_r$), ($H_c$) and squareness ratio ($M_r/M_s$) as particle size decreased. Among the compositions, BCFOCr-5 exhibited the highest $M_r$, reaching approximately 28 times that of pure BFO. However, the maximum $M_s$ was obtained with Ca as the sole dopant, showing an increase of over 3 times compared to pure BFO.

[ Insert Figure 4 around here]

As reported in literature, a way to unwind the helical spin cycloid that governs the antiferromagnetic nature of BFO is through particle size control via doping [36,37]. The highest magnetization among the co-doped samples is seen in BCFOCr-5, where the particle size (61 nm) is beneath the threshold of the spin-cycloid wavelength (62 nm), destabilizing the structure and eliminating the spin-cancellation effect as a result. Beyond the helical structure, other factors are known to have an effect on the magnetic behavior of doped BFO NPs, such as i) increased surface to volume ratio leading to uncompensated Fe spins at the particle surface, forming a ferromagnetic shell with an antiferromagnetic core [38] ii) the presence of oxygen vacancies [39] and iii) increased spin canting due to lattice strain [40]. From **Table 2**, it is evident that increasing Cr concentration led to an increase in lattice strain.

According to Néel's theory, this long-range anti-parallel spin compensation between sublattices may be interrupted at particle surfaces for nanoscale antiferromagnets, deviating from bulk properties [41]. As the particle size decreases, the surface-to-volume ratio increases, which can enhance the net magnetization due to the presence of uncompensated spins on the surface. Park et. al. observed a linear relationship between net magnetization and 1/d for single-domain antiferromagnetic BFO nanoparticles (where d is the particle diameter), and modelled the structure by superposition of an AFM Core with a FM Shell, as shown in **Figure 5(a)** [18]. This corroborates our data of increased magnetization with a decrease in particle size. The formation of this core-shell structure can be further confirmed by the presence of exchange bias, as observed by the shift of the hysteresis loops along the magnetic field axis in **Figure 4**. Exchange bias is a phenomenon commonly present in thin films with FM and AFM layers, where the AFM layer resists rotation of its magnetic configuration and pins the moment of the FM component, leading to asymmetrical hysteresis loops. By modeling the nanoparticles using the core-shell model, a phenomenon similar to the exchange bias effect occurs at the FM-AFM interface, leading to the shift in the hysteresis curves [42]. The hysteresis shift ($H_{EB}$) was calculated by $H_{EB}=(H_{c1}+H_{c2})/2$ where the right and left coercive fields at zero magnetization are represented by $H_{c1}$ and $H_{c2}$, respectively.

[Insert Figure 5 around here]

A possible explanation for the deviation of BCFO from the Magnetization ~1/d trend could be due to increased ferromagnetic behavior from uncompensated spins resulting from the creation of oxygen vacancies upon $Ca^{2+}$ substitution [43]. As shown by the Kröger-Vink notation in **Equation 1**, the substitution of trivalent $Bi^{+3}$ with divalent $Ca^{2+}$ leads to a charge imbalance, which is then compensated either by the creation of oxygen vacancies or through the oxidation of $Fe^{3+}$ to $Fe^{4+}$, maintaining charge neutrality [44]. XPS analysis revealed that upon Ca doping, charge neutrality in the system was maintained primarily through the formation of oxygen vacancies, rather than by changes in the oxidation state of Fe. $Fe^{3+}$ was found to be the dominant oxidation state, consistent with findings reported by Chauhan et al. [45]. This leads to the maximization of $M_s$ in BCFO. Previous researchers have also reported an increase in magnetization behavior due to the formation of oxygen vacancies [46] and this could likely account for the BCFO NPs having the highest magnetization despite the particle size effect. The coercive field is determined by the formula $H_c=(H_{c1}-H_{c2})/2$. The increase in $H_c$ with increasing dopant concentration could also be attributed to increased inter-particle boundary area upon size reduction. This restricts domain wall motion through pinning and blocking, which in turn increases coercivity and remnant magnetization. [47]. Maximum $H_c$ was obtained for BCFOCr-5, possessing the smallest particle size. Increased coercivity could also be attributed to increased lattice strain upon $Cr^{3+}$ substitution [48]. Along with increased coercivity, the squareness ratio also improved, a critical parameter for incorporating BFO in magnetic memory-based device applications.

**Table 3**

Magnetic Properties of nanoparticles

| Samples | Coercive field, $H_c$ (Oe) | Remnant Magnetization, $M_r$ (emu/gm) | Saturation Magnetization, $M_s$ (emu/gm) | Squareness, $M_r/M_s$ | Exchange Bias Shift, $H_{EB}$ (oe) | Particle Size, d (nm) | Néel Temperature, K ($T_n$) |
|---|---|---|---|---|---|---|---|
| BFO | 24.5 | 0.02 | 1.075 | 0.02 | 0.5 | 83 | 645 |
| BCFO | 55 | 0.48 | 3.70 | 0.13 | 10 | 74 | 743 |
| BCFOCr-1 | 95 | 0.19 | 1.59 | 0.12 | 91 | 71 | - |
| BCFOCr-3 | 81.5 | 0.18 | 1.98 | 0.10 | 37 | 67 | - |
| BCFOCr-5 | 115 | 0.56 | 3.30 | 0.17 | 10 | 61 | 610 |

The magnetic response of 3 samples (BFO, BCFO, BCFOCr-5) were recorded over a temperature range from 300K to 900K. The onset of paramagnetic behavior signifies the Néel Temperature of antiferromagnets, as shown in **Figure 5(b)**. $T_N$ of pure BFO was found to be around ~645K, which is in alignment with the values obtained from literature [49]. However, $Ca^{2+}$ enhances the magnetic transition temperature to higher values while co-doping with Ca and Cr decreases it, as observed by the $T_N$ of our BCFO and BCFOCr-5 samples being ~743K and 610K respectively. The effect of $Ca^{2+}$ doping on magnetic ordering temperature has been studied previously [50-53]. Catalan et al. noted that replacing $Bi^{3+}$ with $Ca^{2+}$, an element with a smaller ionic radius, produced a chemical pressure that led to reduction of both the volume and the rhombohedral distortion of the unit cell, resulting in increased Néel temperatures [51]. Tahir et al. and Kamynin et al. also noted a similar shift of magnetic ordering temperatures to higher values upon incorporation of a dopant with a smaller ionic radius and attributed it to the decrease in the unit cell volume, leading to changes in the energy of the exchange interactions and requiring greater thermal energy for magnetic disordering [52,53].

The lower Néel temperature ($T_n$) observed for BCFOCr-5 could be attributed to its smaller particle size, consistent with studies reporting a decrease in $T_N$ with reduction in crystallite size [54]. Selbach et al. [55] reported a correlation between the number of antiferromagnetic (AFM) interactions in a specific size region to the magnetic ordering temperature, which aligns with our findings. The weakened AFM interactions in BCFOCr-5 are likely due to a higher fraction of uncompensated surface spins, reducing the overall antiferromagnetic coupling. At temperatures above Néel point, all three NP compositions showed rapid demanganization as it entered a paramagnetic phase where the magnetic moments are randomly ordered [56]. This is in accordance to the Curie Weiss law :

$$\chi = \frac{C}{T + \theta} \tag{7}$$

*3.4 Ferroelectric Analysis*

The ferroelectric properties of BFO, BCFO, and BCFOCr-5 NPs were evaluated through P-E hysteresis loops recorded at room temperature. Measurements were carried out under an electric field ranging from -10 to +10 kV/cm at a frequency of 1 Hz. The open-ended nature of the curves suggests the presence of oxygen vacancies or secondary phases which lead to higher amounts of leakage current. Matin et al., also reported oxygen vacancies as a key factor in limiting the ferroelectric properties of BFO [57]. Although the doped NPs show a uniform polarization response, they exhibit significant leakage current, which could be attributed to the introduction of oxygen vacancies upon $Ca^{2+}$ substitution, as supported by the XPS data. Another reason for high leakage current could be due to presence of defects such as porosity or surface cracks upon pressing the NPs into pellets. These extrinsic defects, rather than intrinsic material properties, significantly influence the electrical behavior by acting as conductive pathways for charge carriers. That said, introduction of $Ca^{2+}$ and $Cr^{3+}$ doped BFO NPs show improved remnant polarization in BCFO and BCFOCr-5 as shown in **Figure 6.**

[Insert Figure 6 here]

*3.5 XPS Analysis*

To determine the surface chemical composition, valence states of the host (Bi, Fe, O) and dopant (Ca, Cr) elements and to quantify the amount of oxygen vacancies present in the NPs, X-ray photoelectron spectroscopy (XPS) was performed. **Figure 7** displays the XPS survey spectra, where presence of Bi, Fe, and O are confirmed. The presence of characteristic Ca peaks and both Ca and Cr peaks in BCFO and BCFOCr-5 respectively indicate successful dopant incorporation in the NPs. The presence of carbon (C) in all the samples can be attributed to unavoidable surface contamination upon exposure to air, where carbon-based contaminants, such as hydrocarbons, adsorb onto the sample surface [58]. **Figure 8 (a-i)** illustrates the high-resolution core spectra of BFO, BCFO, and BCFOCr-5 samples, showing the binding energy regions for Bi 4f (a-c), O 1s (d-f), and Fe 2p (g-i).

[ Insert Figure 7 and 8 around here]

For the Bi 4f core spectra, the NPs demonstrate doublet peaks, which are likely signals from the Bi-O bonds. For BFO, these doublet peaks are observed at 159.16eV and 164.46eV, corresponding to the binding energy of Bi $4f_{7/2}$ and Bi $4f_{5/2}$ respectively [34]. However, these peaks were slightly shifted to higher energies for BCFO, appearing at 159.26 eV and 164.48 eV. Following this trend, the peaks are further shifted to higher binding energies at 159.46 and 164.76 eV respectively for BCFOCr-5. This shift can be attributed to the ionicity fraction of the Bi-O and Ca-O bonds, which was determined using Pauling's expression for ionicity fraction ($f_{ionic}$) [59]:

$$f_{ionic} = 1 - e^{-\frac{1}{4}(x_{nm}-x_m)^2} \qquad (8)$$

Here $x_{nm}$ and $x_m$ represents the electronegativity values of the nonmetallic element (O) and metallic element (Bi, Cr) respectively. **Table 4** lists the ionicity fraction of the bonds along with the electronegativities of their respective elements [59]. $f_{ionic}$ of Ca-O bond being higher than the Bi-O bond indicates stronger ionic characteristics, resulting in a higher binding energy. **Figure 8(d-f)** displays the XPS spectra of the O 1s core level for BFO, BCFO and BCFOCr-5 samples. For pure BFO, the spectra were deconvoluted into two peaks at 529.78 eV and 531.68 eV, corresponding to lattice oxygen ($O_L$) and oxygen vacancies ($O_V$), respectively [60]. Formation of oxygen vacancies in BFO could be attributed to factors such as volatilization of Bi during the annealing [61]. The Kröger-Vink notation for these phenomena are given below:

$$Bi_{Bi}^x + 3O_O^x \rightarrow 2V_{Bi}''' + 3V_O^{\bullet\bullet} + Bi_2O_3(\uparrow) \qquad (9)$$

$$O_O^x + 2Fe_{Fe}^x \rightarrow 2Fe_{Fe}' + V_O^{\bullet\bullet} + \frac{1}{2}O_2(\uparrow) \qquad (10)$$

For BCFO, the $O_L$ and $O_v$ peaks shifted slightly to higher binding energies, observed at 529.88 eV and 531.78 eV, respectively, which is attributed to the stronger bond between Ca and O. Peak area ratio of oxygen vacancies to oxygen lattice, $(O_V)/(O_L)$, increased from 0.87 in BFO to 0.93 in BCFO, indicating that Ca doping enhances oxygen vacancy formation, according to **Equation 1**. This equation shows that the incorporation of $Ca^{2+}$ into the $Bi^{3+}$ site introduces a charge imbalance, which can be neutralized by either forming oxygen vacancies or through oxidation of $Fe^{3+}$ to $Fe^{4+}$ [23]. The absence of $Fe^{4+}$ peaks suggests that charge compensation during $Ca^{2+}$ substitution occurred primarily through the creation of oxygen vacancies rather than the oxidation. In the BCFOCr-5 sample, the $O_L$ and $O_V$ peaks were further shifted to 530.18 eV and 532.18 eV, with an $(O_V/O_L)$ area ratio of 0.90, demonstrating a higher oxygen vacancy content compared to pure BFO yet lower than BCFO.

**Table 4:**

Electronegativity and ionicity fractions values of Bi, Fe, Ca, Cr and O Bonds.

| Bond (Element 1- Element 2) | Electronegativity (Element 1) | Electronegativity (Element 2) | Ionicity Fraction ($f_{ionic}$) |
|---|---|---|---|
| Bi-O | 1.9 | 3.5 | 0.47 |
| Fe-O | 1.8 | | 0.51 |
| Ca-O | 1.0 | | 0.79 |
| Cr-O | 1.6 | | 0.59 |

**Figure 8(g-i)** shows Fe 2p XPS core spectra for BFO, BCFO, and BCFOCr-5 NPs. Characteristic doublet and shake-up satellite peaks were observed in all cases. The peaks for the $Fe^{3+}$ were observed at 710.86 eV for BFO, 710.88 eV for BCFO, and 711.36 eV for BCFOCr-5. According to literature, the characteristic satellite peak for $Fe^{3+}$ appears at around 8eV higher the main $2p_{3/2}$ peak [60] which is in good agreement with our experimental data, confirming the dominance of the $Fe^{3+}$ oxidation state. Quantitative analysis of peak areas indicated $Fe^{3+}$:$Fe^{2+}$ ratios of 1:0.164 for BFO, 1:0.150 for BCFO, and 1:0.194 for BCFOCr-5, with BCFO showing the highest $Fe^{3+}$ content while BCFOCr-5 exhibiting the highest $Fe^{2+}$ content. While $Fe^{3+}$ dominance is maintained in all NPs, due to increased local strain and defect density, $Cr^{3+}$ substitution may destabilize some amounts of $Fe^{3+}$ and promote $Fe^{2+}$ formation, as observed in BCFOCr-5 [62].

## 4   Conclusions

This study explored the effects of A-site $Ca^{2+}$ and B-site $Cr^{3+}$ co-substitution in $BiFeO_3$ (BFO) through the synthesis of $Bi_{0.97}Ca_{0.03}Cr_xFe_{1-x}O_3$ compositions, addressing a key gap in the literature. Magnetic hysteresis loop measurements revealed a significant enhancement in saturation ($M_s$) and remanent magnetization ($M_r$), driven by spin cycloid suppression. The presence of an exchange bias effect confirmed an underlying FM-AFM core-shell interaction, further stabilizing the magnetic behavior. Additionally, Néel temperature ($T_N$) increased upon divalent ion substitution, indicating enhanced magnetic stability. XPS analysis confirmed the formation of oxygen vacancies and the predominance of $Fe^{3+}$ oxidation states across all compositions. Ferroelectric polarization exhibited improved remanent polarization in doped samples, despite leakage characteristics linked to oxygen vacancies and extrinsic microstructural defects. The observed magnetic property enhancements and exchange bias are particularly interesting, as they open pathways for BFO's potential integration into a range of magneto-electric memory devices (FeRAM, MeRAM) , spintronics and quantum computing. Future work could explore site engineering strategies to further enhance magnetoelectric coupling, minimize leakage currents, and optimize band gap tuning, ultimately broadening the practical applications of co-substituted BFO in next generation magneto-electric and opto-electronic applications.

**Declaration of competing interest**

The authors declare that they have no known competing financial interests or personal relationships that could have appeared to influence the work reported in this paper.

**Data availability**

The data that support the findings of this study are available from the corresponding author upon reasonable request.

**Acknowledgements**

The authors would like to acknowledge the department of Nanomaterials and Ceramic Engineering for their assistance with nanoparticle synthesis and characterization.

**List of Figures**

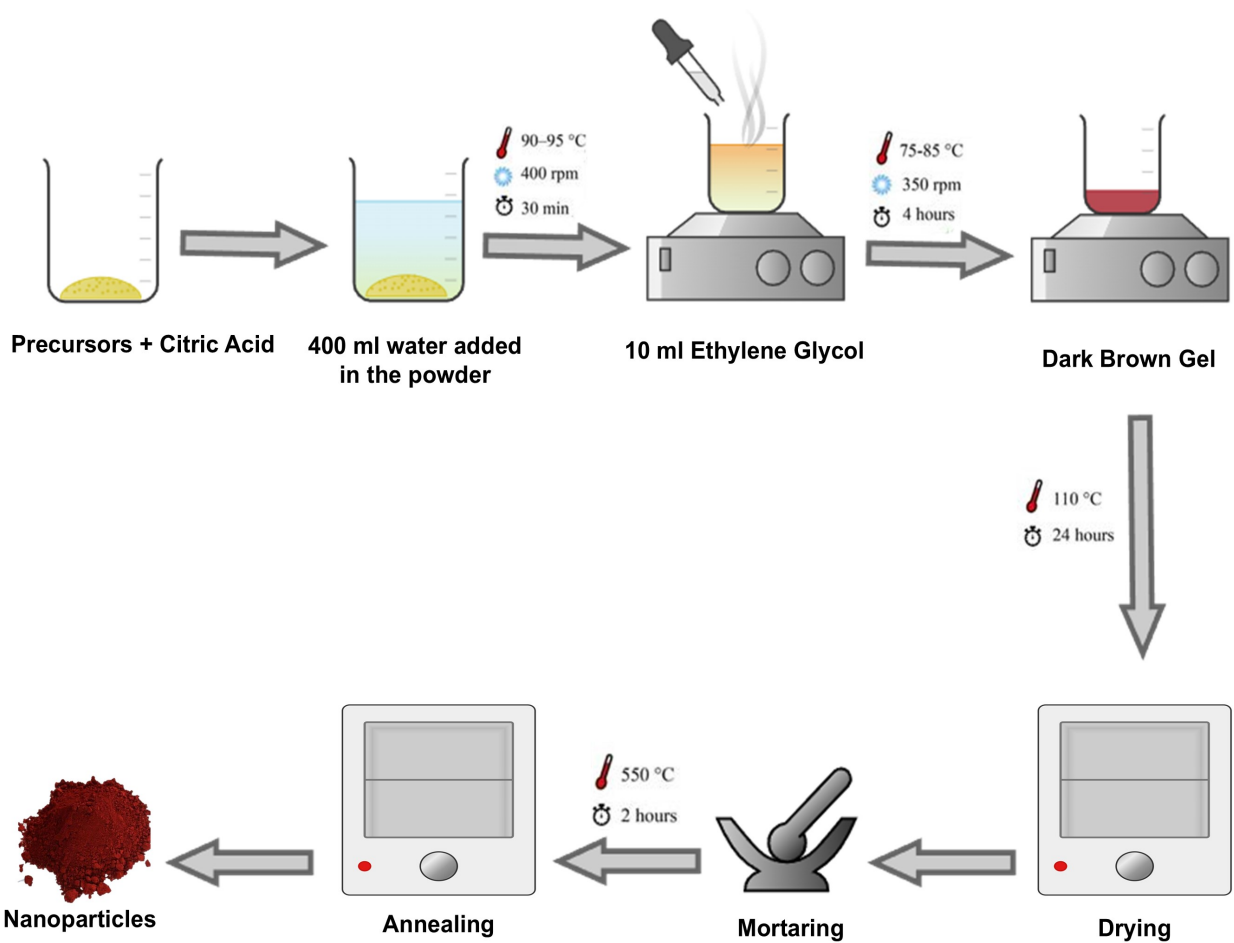

**Figure 1:** Schematic diagram of the nanoparticle synthesis route.

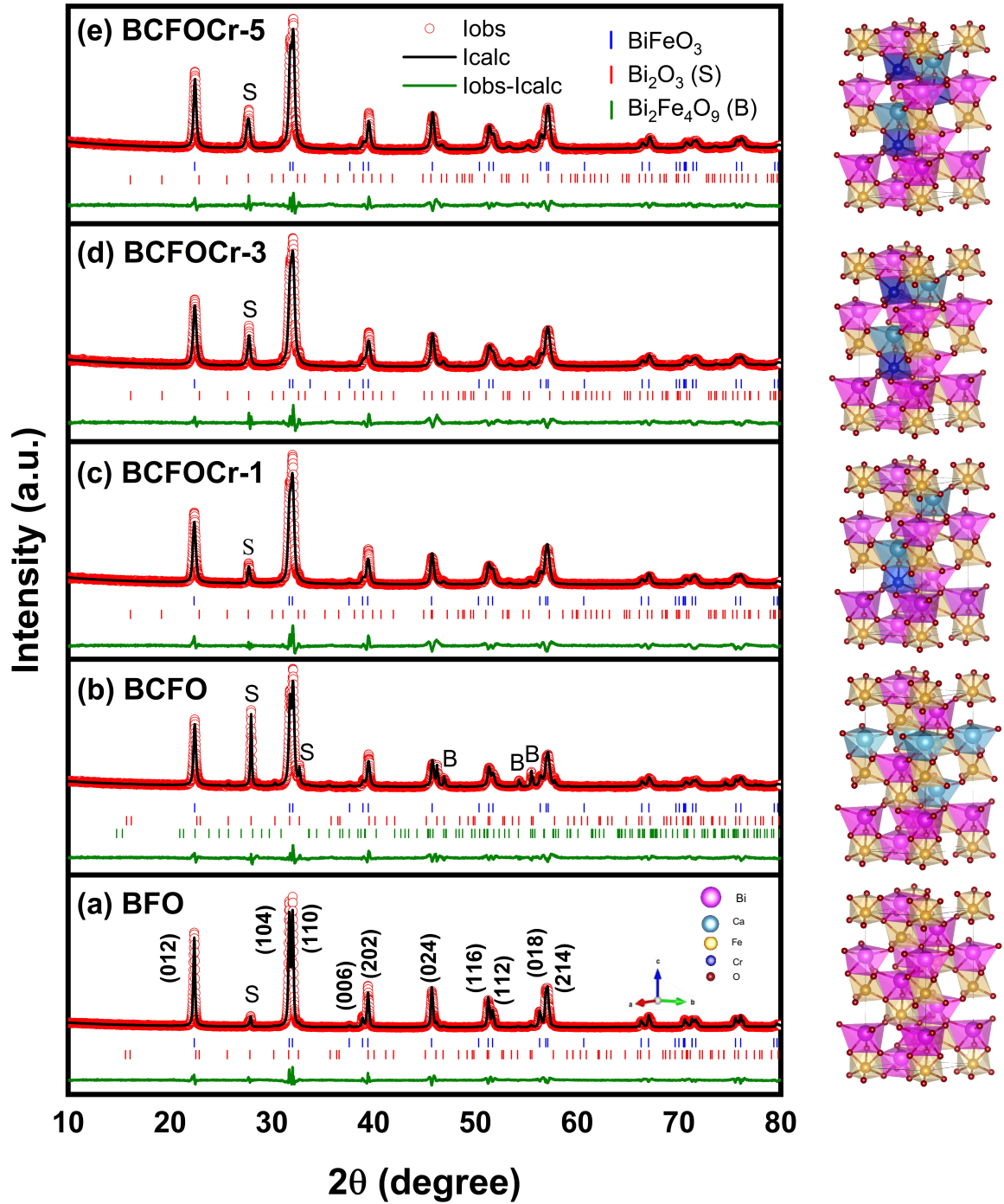

**Figure 2:** Rietveld analysis of room temperature XRD patterns of (a) BFO, (b) BCFO, (c) BCFOCr-1, (d) BCFOCr-3, (e) BCFOCr-5 nanoparticles along with their corresponding crystal structures.

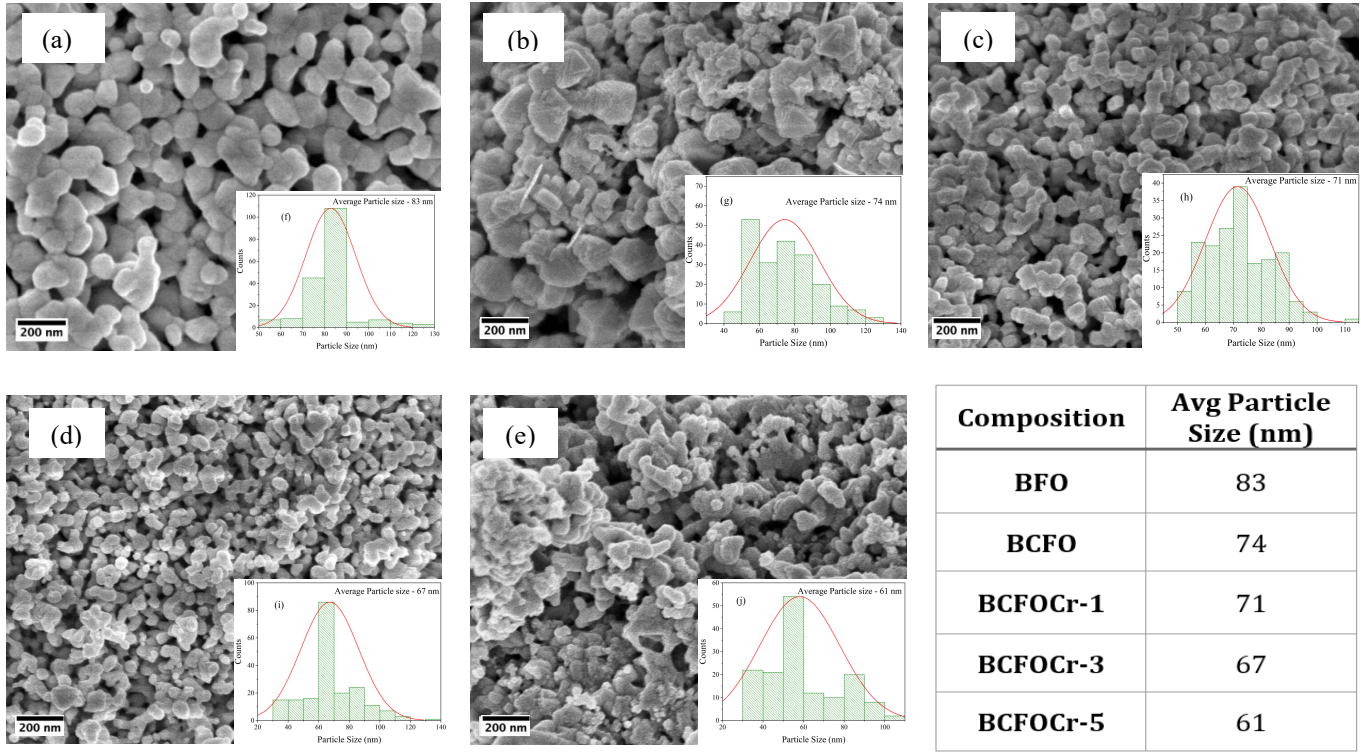

**Figure 3:** FESEM micrographs of pure and doped BFO (a-e) and their average particle sizes.

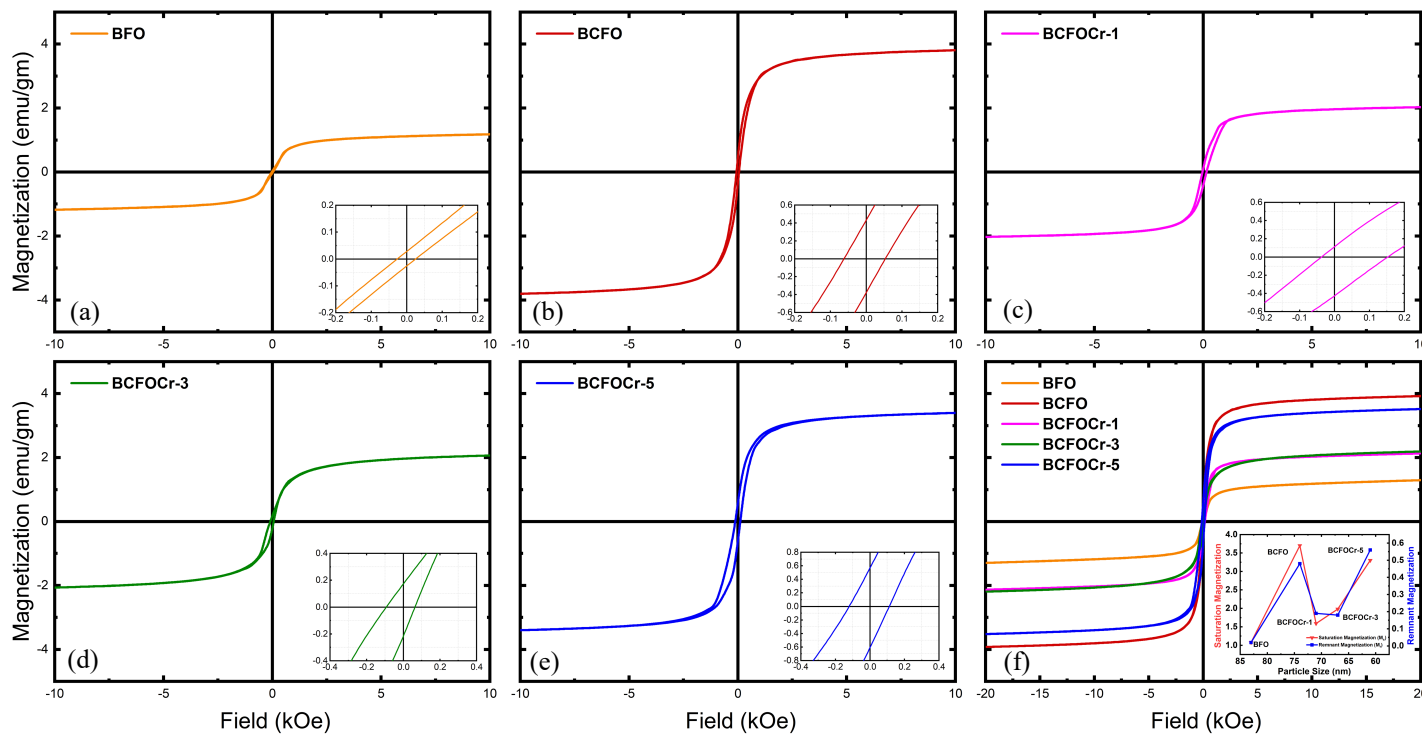

**Figure 4:** RT M-H hysteresis loops of nanoparticles measured at room temperature. Inset : Zoomed in M-H curves to determine coercivity and remnant magnetization (f) Combined Hysteresis Loops. Inset : Correlation between particle size, $M_s$ and $M_r$.

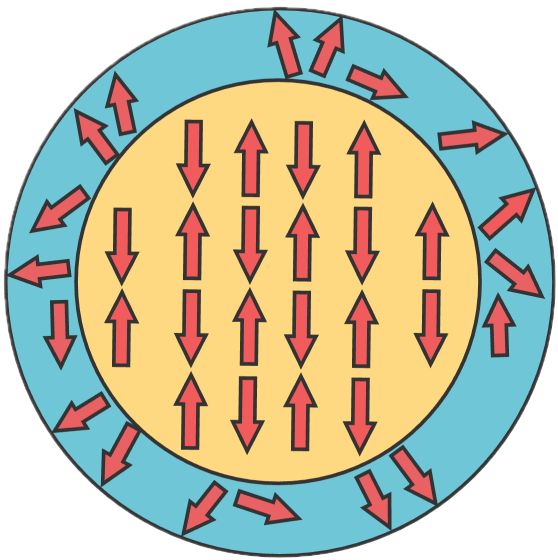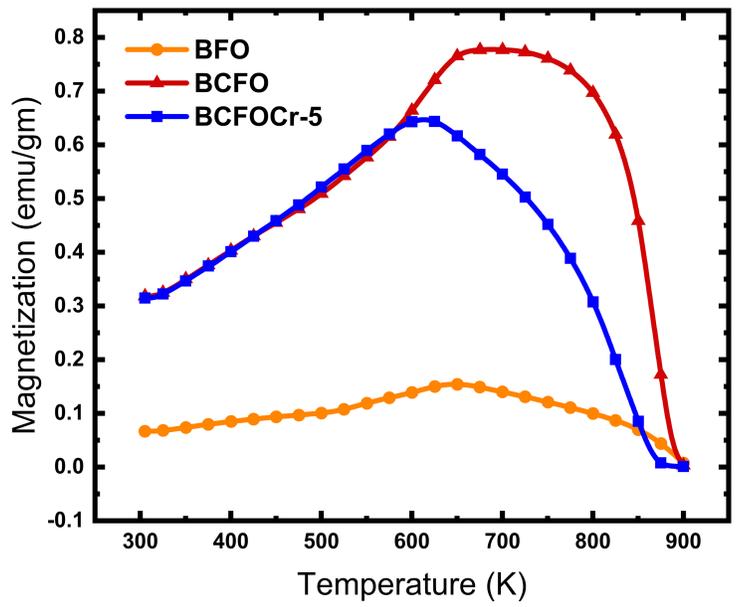

**Figure 5:** (a) Schematic modeling of AFM Core-FM Shell structure. (b) Magnetization vs Temperature tests conducted for NPs.

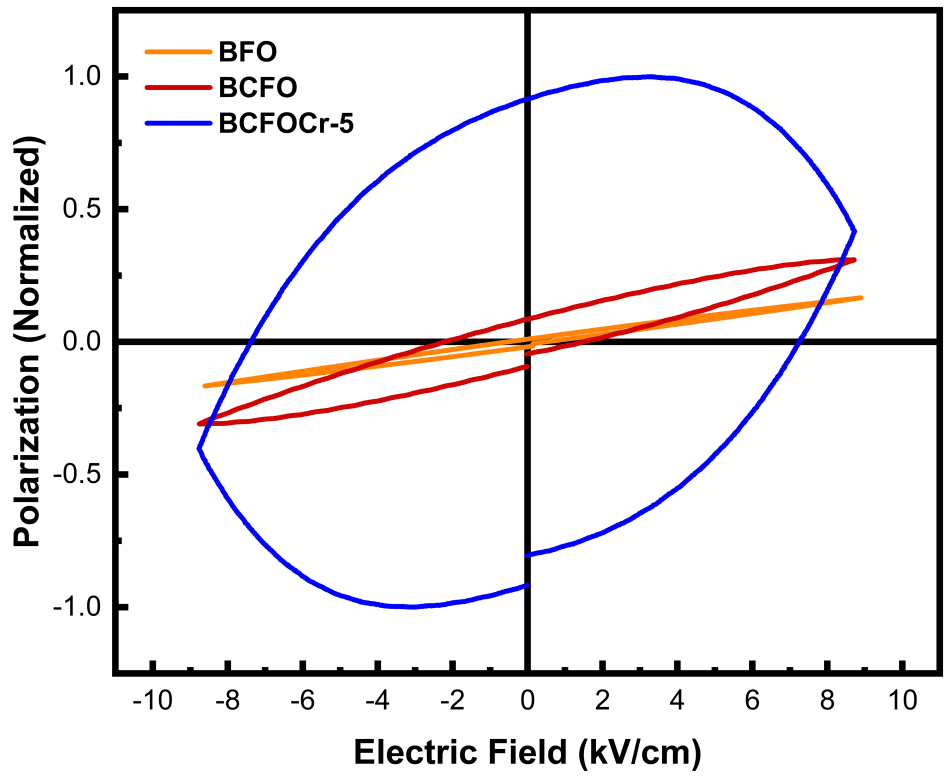

**Figure 6:** Combined P-E loops corresponding to BFO, BCFO and BCFOCr-5 (scaled in terms of $P_{max}$ of BCFOCr-5).

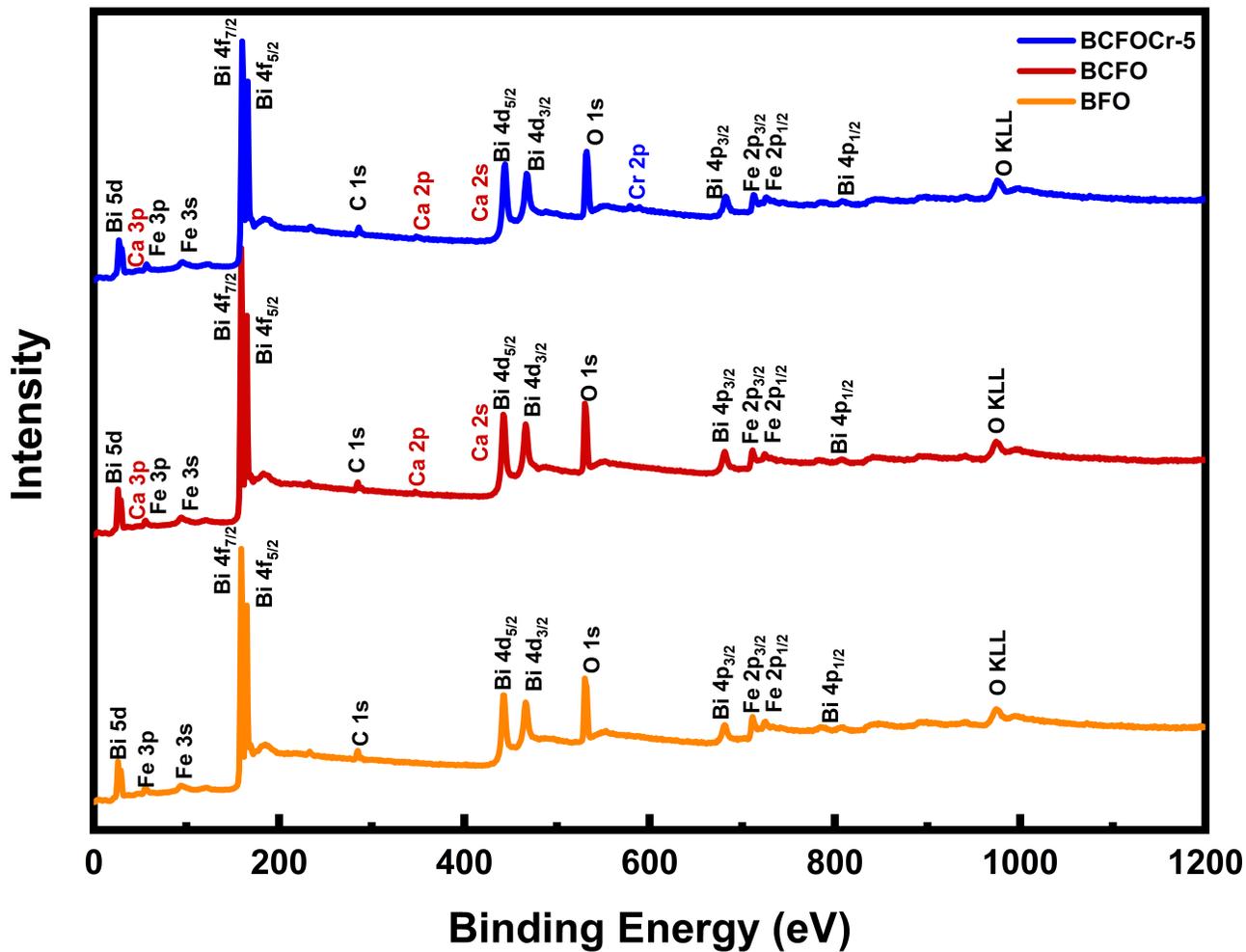

**Figure 7**: XPS survey spectra of BFO, BCFO, and BCFOCr-5 samples.

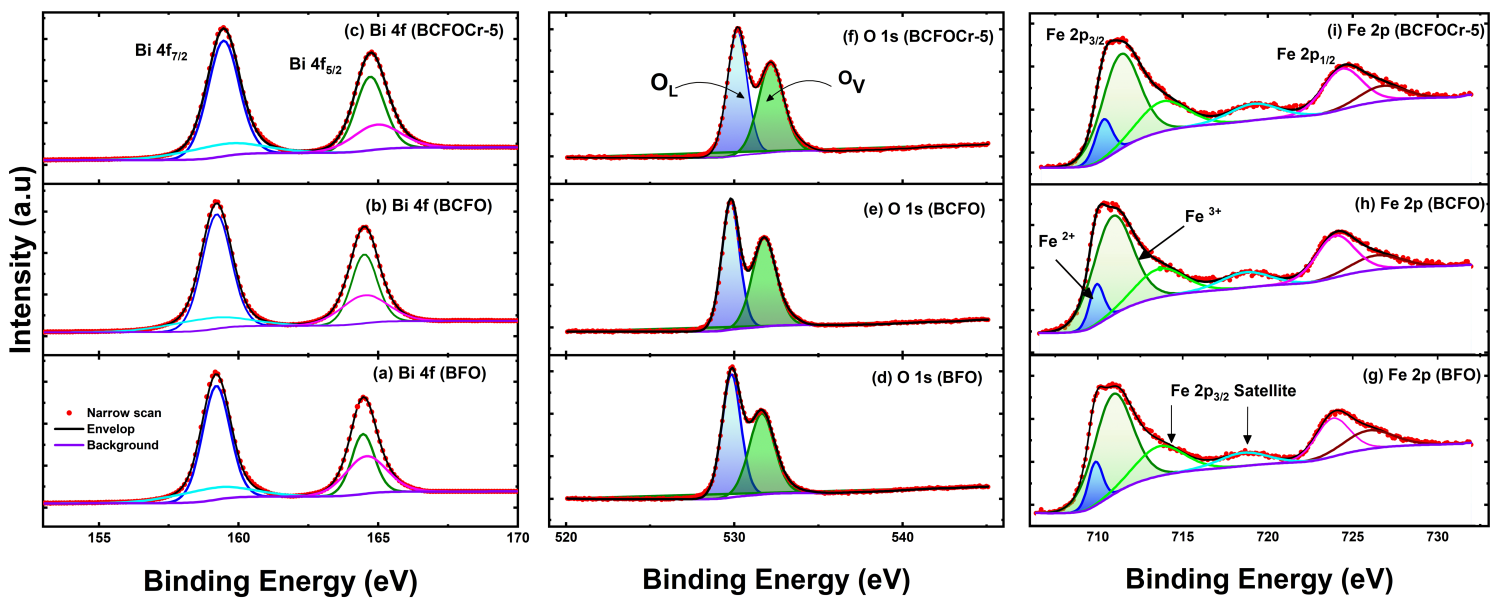

**Figure 8:** XPS core level spectra for Bi 4f, O 1s, and Fe 2p in BFO, BCFO, and BCFOCr-5 samples. (a)–(c) Bi 4f spectra showing peaks corresponding to Bi $4f_{7/2}$ and Bi $4f_{5/2}$ states. (d)–(f) O 1s spectra displaying lattice oxygen ($O_L$) and oxygen vacancy ($O_V$) contribution. (g)–(i) Fe 2p spectra, highlighting $Fe^{2+}$, $Fe^{3+}$, and satellite peaks, indicating mixed Fe oxidation states across the samples.